# Gap-type dark localized modes in a Bose–Einstein condensate with optical lattices

Liangwei Zeng[a,b] and Jianhua Zeng[a,b,*]
[a]Chinese Academy of Sciences, Xi'an Institute of Optics and Precision Mechanics, State Key Laboratory of Transient Optics and Photonics, Xi'an, China
[b]University of Chinese Academy of Sciences, Beijing, China

**Abstract.** Bose–Einstein condensate (BEC) exhibits a variety of fascinating and unexpected macroscopic phenomena, and has attracted sustained attention in recent years—particularly in the field of solitons and associated nonlinear phenomena. Meanwhile, optical lattices have emerged as a versatile toolbox for understanding the properties and controlling the dynamics of BEC, among which the realization of bright gap solitons is an iconic result. However, the dark gap solitons are still experimentally unproven, and their properties in more than one dimension remain unknown. In light of this, we describe, numerically and theoretically, the formation and stability properties of gap-type dark localized modes in the context of ultracold atoms trapped in optical lattices. Two kinds of stable dark localized modes—gap solitons and soliton clusters—are predicted in both the one- and two-dimensional geometries. The vortical counterparts of both modes are also constructed in two dimensions. A unique feature is the existence of a nonlinear Bloch-wave background on which all above gap modes are situated. By employing linear-stability analysis and direct simulations, stability regions of the predicted modes are obtained. Our results offer the possibility of observing dark gap localized structures with cutting-edge techniques in ultracold atoms experiments and beyond, including in optics with photonic crystals and lattices.

## 1 Introduction

Bose–Einstein condensate (BEC) consists of interacting and ideal dilute Bose gases cooled down to a very low temperature (i.e., around absolute zero) and is one of the most famous examples of the macroscopic quantum phenomena that have attracted increasing interest in past decades. Because of being equipped with intrinsic nonlinear effect arising from atom–atom collisions, the BEC (and more broadly, ultracold atoms) is an innately nonlinear medium in which there are many emergent nonlinear phenomena, such as matter–wave four-wave mixing, bright and dark solitons, vortices and vortex lattices, and dynamic instabilities.[1] In addition to basic research interests, BEC also provides useful applications in cold atom interferometry, atom lasers, and optical atomic clocks, with unprecedented precision and stability, and, more excitingly, quantum information processing.[1–7] Recently, the study of BEC has maintained a good momentum of development and expansion, attracting increasing attention in various fields, which has manifested its emergence in solid-state matter and its BEC-like phenomenon caused by elementary excitations in solids (bosonic quasiparticles) in conditions of thermodynamic equilibrium;[8,9] the latter can also occur under nonequilibrium conditions, and thus has been dubbed nonequilibrium Bose–Einstein-like condensation.[8,9]

The coherent interaction between diffraction and nonlinearity can reach a balance so as to form solitons. The bright solitons and dark ones are acknowledged to exist in self-focusing and self-defocusing nonlinearity, respectively, with negative and positive atom scattering factors (which determine the strength of nonlinearity) in BEC.[1–3] They are exact analytical solutions of the underlying one-dimensional (1-D) physical model. Note that within the mean-field theory, the dynamics of BEC are usually described by a single-particle Schrödinger-like equation—the proverbial Gross–Pitaevskii (nonlinear Schrödinger) equation. As fundamental excitations of the nonlinear dynamical equation, dark solitons, which feature a localized dip on the condensate density background and are accompanied by a phase jump in the localized center (where the density is zero), can be dynamically stable.[1] Both bright solitons[10–13] and dark ones[14–24] have been created experimentally in atomic condensates with properly chosen elements of atoms. In particular, dark solitons in cigar-shaped BECs of $^{87}$Rb[25] and $^{23}$Na[26] have been created in the first two experiments[25,26] using the magnetic trap to bind the repulsive condensates (the condensates with repulsive interatomic interaction) onto which a steep spatial phase distribution (atomic density gradient) was written by the phase-imprinting technique.

*Address all correspondence to Jianhua Zeng, E-mail: zengjh@opt.ac.cn

Advances in laser technology have enabled the generation of optical lattices formed by coherent interference of two counter-propagating laser beams, which affords significant opportunities for studying lots of fascinating physical phenomena in atomic, molecular, optical, and quantum physics, and particularly for simulating idealized quantum many-body systems (and problems) that are very relevant to the condensed-matter physics community.[2–7] Ultracold atoms placed in optical lattices provide a clear, easy-to-implement, and precisely controlled test bed, which is advantageous over the crystalline lattices in solid-state physics, to study diverse interesting physical properties and rich nonlinear dynamics.[3,6,7] One of the most noteworthy effects is the manipulation to continuous quantum phase transition from a superfluid to the Mott insulator phase (and vice versa) caused by simply varying the depth of the periodic potential.[27,28] The observation of this effect initiated the studies of strong-correlation physics with ultracold atomic gases as an interacting many-body system.[4,5] Particularly exciting is the possibility of investigating a type of bright atomic solitons—matter–wave gap solitons—which exist inside the finite spectral gaps of the underlying linear Bloch-wave spectrum.[2,3,29] Contrary to the above-mentioned traditional thinking, such a localized state can exist under repulsive atomic interactions (defocusing nonlinearity) and be balanced with negative effective dispersion induced by the band edge effect of the optical lattices.[30]

So far, much attention has been focused on the studies of bright solitons, but their dark counterparts[31] do not get the same spotlight (e.g., see Refs. [32–35] and references therein). In particular, the nonlinear dynamical mechanism and properties of dark gap solitons[36–38] and vortical ones[39] supported by optical lattices are not well understood. Furthermore, in two-dimensional (2-D) geometry, as far as we know, the existing possibility of dark matter–wave gap solitons (and vortices) and their veiled properties are yet to be disclosed. It is therefore the motivation of this work to survey them systematically on a theoretical level. We show that placing the BEC into optical lattices can result in the generation of two kinds of dark localized modes—gap solitons and soliton clusters—in both 1-D and 2-D geometric spatial coordinates. It should be pointed out that the dark gap soliton clusters associated with the ground-state nonlinear Bloch waves, which are also a typical feature of all the other gap-type dark modes, are a new kind of localized mode, expanding and perfecting the soliton family—with an emphasis on bright gap waves that were predicted and observed recently in similar physical scenarios with periodic potentials.

Theoretical descriptions, combined with rigorous mathematical solutions of the Gross–Pitaevskii equation repulsive interatomic interaction (self-defocusing Kerr nonlinearity), prove that the so-found dark gap localized modes, supported by a perfect optical lattice, can be dynamically stable extending to the second bandgap (BG) (of the corresponding linear spectrum) in quasi-1-D coordinate. In 2-D cases, however, the situation becomes more complicated, since the gap-type dark structures cannot exist in any perfect optical lattice. To this end, we adopted the defect engineering commonly used in semiconductor technology to introduce single or multiple defects,[40–44] thus forming defective optical structures, inside of which the 2-D dark gap structures as defect modes may be stationed. We show that only this way can we generate stable dark modes. In addition to the dark gap solitons and soliton clusters, their vortical counterparts—dark gap vortices and vortex clusters—both rest on induced bright defects, and can be stable solutions with topological charges $m = 1$ and $m \leq 2$, respectively. Stability regions of all the predicted solutions (both 1-D and 2-D) are corroborated in their respective parameter spaces relying on linear-stability analysis and direct simulations.

This paper is organized as follows: in Sec. 2, we introduce the theoretical model and give its numerical methods, such as Newton's iteration for searching stationary solutions, and stability testing methods based on the linear-stability analysis and direct numerical simulation; in Sec. 3, we report the numerical results for both 1-D and 2-D gap-type dark localized modes, respectively; in Secs. 3.1 and 3.2, including dark gap solitons and soliton clusters for both dimensions and the vortical counterparts for 2-D cases, we describe physical explanations of nonlinear localizations and the self-trapping effect of the coherent matter–wave dark modes; finally, a summary and suggestions for potential future research directions are given in Sec. 4.

## 2 Model and Numerical Method

The physical model that describes the dynamical evolution of matter waves of a BEC trapped in optical lattices can be described in the framework of the Gross-Pitaevskii equation,[1–3] with the wave function $U$:

$$i\hbar \frac{\partial U}{\partial \tau} = -\frac{\hbar^2}{2m}\nabla^2 U + \widehat{V_{\mathrm{OL}}}(r)U + \frac{4\pi\hbar^2 a_s}{m}|U|^2 U, \quad (1)$$

where $\hbar$ is Planck's constant, $m$ is the mass of the atom, $\tau$ denotes the evolutional time, $\widehat{V_{\mathrm{OL}}}$ is the linear trapping potential (an optical periodic potential representing an optical lattice whose structure may be controlled in experiments using the counter-propagating laser beams), Laplacian $\nabla^2 = \partial_X^2$ and $\nabla^2 = \partial_X^2 + \partial_Y^2$ for the atomic BEC media of dimension $D = 1$ and 2, respectively, and the parameter $a_s$ corresponds to the s-wave scattering length that characterizes the coupling strength of two-body interaction. For convenience of discussion, Eq. (1) is usually scaled in a dimensionless form:

$$i\frac{\partial \psi}{\partial t} = -\frac{1}{2}\nabla^2 \psi + V_{\mathrm{OL}}(r)\psi + g|\psi|^2 \psi. \quad (2)$$

The normalization was made by introducing the changes of variables $t = \tau/\tau_0, x/y = 2X/Y, \psi = U/\sqrt{\pi a_{s0}}$, $V_{OL} = m/(4\hbar^2)\widehat{V_{\mathrm{OL}}}$, and by choosing $\tau_0 = m/4\hbar$, and $a_s = |a_{s0}|g$, with $|a_{s0}|$ being the module of the s-wave scattering length and $g$ being the dimensionless nonlinear strength. Here, $g > 0$ corresponds to the strength of defocusing Kerr (or cubic) nonlinearity arising from repulsive interatomic interactions.

Note that this equation also applies to nonlinear optics, with $\psi$ and $t$ being replaced by the field amplitude $E$ and propagation distance $z$, respectively. For the BEC of interest to us, its number of ultracold atoms $N$ (norm) is defined as $N = \iint |\psi|^2 dx dy$.

At certain real chemical potential $\mu$ (in optics, $\mu$ is replaced by propagation constant $-b$), the stationary solutions of Eq. (2) can be written as $\psi = \varphi \exp(-i\mu t)$, with stationary wave function $\varphi$ obeying:

$$\mu \phi = -\frac{1}{2}\nabla^2 \phi + V_{\mathrm{OL}}\phi + g|\phi|^2\phi. \qquad (3)$$

In a 1-D case, we consider a typically used optical lattice ($V_{\mathrm{OL}}$), which is expressed as

$$V_{\mathrm{OL}} = V_0 \sin^2(x), \qquad (4)$$

where $V_0$ is a constant parameter. In the 2-D case, the expression is

$$V_{\mathrm{OL}} = V_0[\sin^2(x) + \sin^2(y)][1 + V_d(x,y)], \qquad (5)$$

where $V_d$ denotes bright defects of the 2-D optical lattice, as shown in Sec. 3.2. For a 2-D scenario, instead of using the homogeneous optical lattice as mentioned above, since it cannot support stable dark localized modes at all, we take the optical lattice with bright defects, as will be detailed below, which results in the stable gap-type dark modes supported by the introduced defects.

The linear stability analysis of the gap-type dark localized modes is a key issue to determine the dynamics and stability properties of the found localized solutions. In order to do this, we set the perturbed wave function as $\psi = [\phi(r) + p(r)\exp(\lambda t) + q^*(r)\exp(\lambda^* t)]\exp(-i\mu t)$, where $\phi(r)$ is the nondisturbed wave function found from Eq. (2), with $p(r)$ and $q^*(r)$ being small perturbed eigenmodes with eigenvalue $\lambda$. Substituting such expression into Eq. (1) results in the following eigenvalue problems:

$$\begin{cases} i\lambda p = -\frac{1}{2}\nabla^2 p + (V_{\mathrm{OL}} - \mu)p + g\phi(2\phi^* p + \phi q), \\ i\lambda q = +\frac{1}{2}\nabla^2 q + (\mu - V_{\mathrm{OL}})q - g\phi^*(2\phi q + \phi^* p). \end{cases} \qquad (6)$$

According to the above eigenvalue equations, the perturbed dark soliton solutions are stable as long as the real part of all the relevant eigenvalues ($\lambda$) is null, that is, $\mathrm{Re}(\lambda) = 0$.

In Sec. 3, we focus our interest on presenting numerical results of the various gap-type dark localized solutions such as 1-D and 2-D fundamental dark gap solitons and soliton clusters, as well as the ones carrying vortex charge in dimension $D = 2$. First, their stationary solutions were obtained by solving Eq. (2) via Newton's method, and then their stability was examined using the linear-stability analysis and reviewed by direct numerical simulations of the perturbed solutions via Eq. (1). The numerical experiments of the latter two were done based on the widely used finite-difference method.

## 3 Numerical Results

### 3.1 One-dimensional Matter–Wave Dark Gap Localized Structures

To corroborate our theoretical predictions of the above-mentioned matter–wave dark gap localized modes, we show here the relevant numerical results. The 1-D physical setting with the optical potential expression [Eq. (4)] is first considered. To do so, one should portray the relevant band structure of the underlying linear equation (by setting $g = 0$), which has been done and readers can refer to the Supplementary Material, where optical lattices with shallow, moderate, and strong modulations are included, by varying the strength $V_0$. For the sake of discussion, we set $V_0 = 3$ for all the 1-D results, and the nonlinear strength $g = 1.5$ is taken throughout the paper.

Depicted in Fig. 1(a) is the relation between norm $N$ and chemical potential $\mu$ for 1-D dark gap solitons supported by the moderately modulated optical lattice, from which we can see the $N$ has a linear increasing relation with $\mu$, when going deeper inside the higher BGs. It is obvious that the slope of $N(\mu)$ in the second BG is slightly larger than its counterpart in the first BG. Meanwhile, it is seen that—similar to their bright counterparts, bright gap solitons—the gap-type dark modes also obey the empirical "anti-Vakhitov–Kolokolov" (anti-VK) stability criterion,[46–48] that is, $dN/d\mu > 0$. Employing the linear-stability analysis based on eigenvalue equations [Eq. (6)], we can obtain the dependence of the maximal real part of eigenvalues $\mathrm{Re}(\lambda)$ on $\mu$ of such gap-type dark localized modes, as shown in Fig. 1(b), from which we observe that such dark localized modes are robustly stable. Exceptionally unstable ones exist only when they are approaching the band edge, resembling those obtained for their bright counterparts.[49] Remarkably, we also find from Fig. 1(b) that the dark gap solitons are more stable in the higher BG (actually, they are almost completely stable in the second BG), contrary to those obtained before (Ref. 49) for bright ones, where the stability region in the second gap shrinks quickly. This fact is consolidated too by the corresponding dynamical evolution of such dark gap modes in real time by solving the physical model Eq. (1); see the examples in the Supplementary Material.

With further insight into typical shapes of such dark gap solutions thus found, as depicted in Figs. 1(c)–1(e), we find that the central dip of these dark solitons gets narrower with an increase of $\mu$, and interestingly, the central dip shrinks more rapidly when they are prepared in the second BG, since inside they experience a much stronger localization induced by the Bragg scattering therein. It is observed from Fig. 1(e) that such a central dip localized at the minimum of the axial optical periodic potential (and thus, the dark gap solitons) may be viewed as a fundamental ground state of the physical system. Furthermore, a unique feature is that the dark gap solitons are always accompanied with a bilateral periodic wave background, which are actually the nonlinear Bloch waves—the nonlinear counterparts of the well-known linear Bloch-wave solutions—of the underlying Eq. (1) for periodic solutions; such dark gap modes thus may be viewed as Bloch-wave modulated dark gap solitons. In other words, these dark gap solitons are always situated on the relevant nonlinear Bloch waves. The associated relation between both types of nonlinear wave solutions is detailed in the Supplementary Material. It should be noted that this feature holds for all the dark gap solutions reported here.

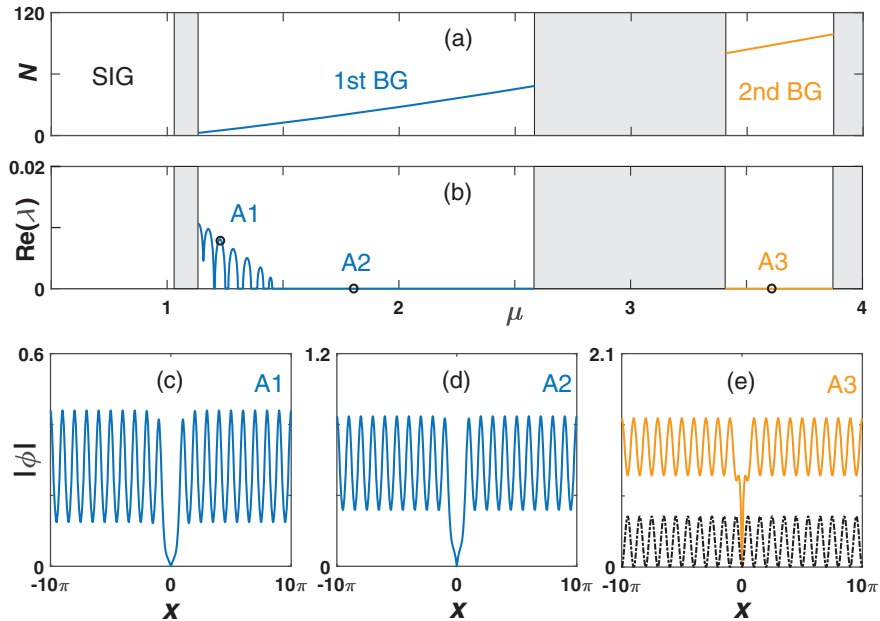

**Fig. 1** (a) Number of atoms ($N$) and (b) maximal real part of eigenvalues versus chemical potential $\mu$ for 1-D matter–wave dark gap solitons found in the model with a 1-D periodic optical potential (optical lattice). The gray areas in this and other figures are the bands of linear spectra. Profiles of 1-D dark gap solitons for three marked circles in panel (b): in first BG with (c) $\mu = 1.23$ and (d) $\mu = 1.8$, and in second BG with (e) $\mu = 3.6$. Here and in Fig. 2, we set $V_0 = 3$, and we set $g = 1.5$ throughout the paper. SIG in panel (a) [and in Figs. 2(a) and 4(a)] denotes the semi-infinite gap. Black dashed line in panel (e) represents the scaled shape of the optical lattice.

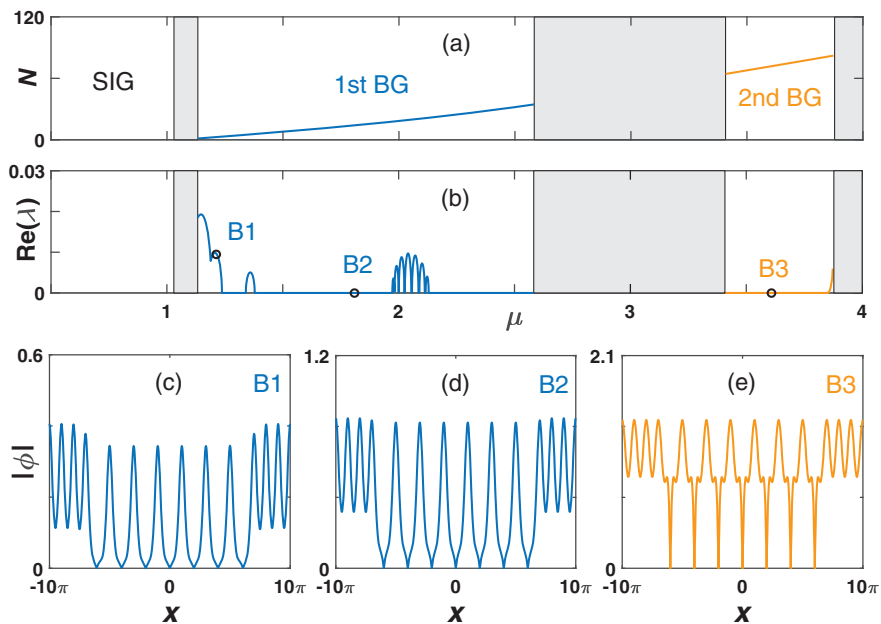

**Fig. 2** The same as in Fig. 1 but for families of 1-D matter–wave dark gap soliton clusters (composed of seven individuals), with which the nonlinear Bloch waves are accompanied. In the bottom panels (c)–(e), the spacing ($\Delta$) between adjacent solitons is $\Delta = 2\pi$, doubling the period of the optical lattice. The chemical potential $\mu = 1.2$ for panel (c); its values for panels (d) and (e) are the same as those in Figs. 1(d) and 1(e), respectively.

Next, we try to construct higher-order nonlinear excitations of 1-D dark gap solitons, which we call dark gap soliton clusters. In reality, arranged solitons are what constitute these soliton clusters—hence the name. The curve $N(\mu)$ for the 1-D dark gap soliton clusters formed by seven individuals (identical dark solitons) is plotted in Fig. 2(a), where the anti-VK stability criterion $dN/d\mu > 0$ remains valid; plotting is also for their stability and instability region, shown as $\text{Re}(\lambda)$ versus $\mu$, within the first two finite BGs in Fig. 2(b). Profiles of such soliton clusters are displayed in the bottom panels [Figs. 2(c)–2(e)], where the spacing ($\Delta$) between adjacent solitons is defined as $\Delta = 2\pi$, which is twice the period of the optical lattice, $\Delta_{\text{latt}} = \pi$. Surely, the dark gap soliton clusters with different soliton spacing ($\Delta$) can exist and be stable too, as long as the requirement $\Delta = n\Delta_{\text{latt}} = n\pi$ (with $n \geq 2$) is met (our simulations verified that soliton clusters are always unstable under the condition $\Delta = \Delta_{\text{latt}} = \pi$). We point out that direct simulations of evolution of localized modes shown in Figs. 2(c)–2(e) are included in the Supplementary Material.

### 3.2 Two-dimensional Matter–Wave Dark Gap Localized Structures

Generating stable high-dimensional dark gap localized structures is a nontrivial issue, owing to the presence of strong localization effects in the respective BGs of the underlying linear spectrum. In fact, things get a little bit complicated because of the unmodified 2-D optical lattice (perfect shape): simply extending the 1-D case to a higher domain does not guarantee the formation of dark gap solitons any more (this conjecture has also been corroborated numerically). It is therefore still an open issue to find gap-type dark localized modes higher than one dimension. To explore a possible way that allows for their existence and that can be realized with general techniques already used in current ultracold atom experiments is a major motivation for this paper.

Can the defected optical potentials (imperfect optical lattices that may break the spatial symmetry) accomplish this mission? It might work, considering the fact that defect engineering is widely used in realizing all-optical integrated circuits from diverse optical periodic potentials, such as photonic crystals and lattices. Based on this method, stable 2-D bright solitons and vortices, in a BEC trapped in optical lattices, emerging as localized defect modes within the gap spectrum region, have been reported. Particularly, Zeng and Malomed[50] demonstrated that one or several dark defects (induced as holes) can support various bright gap-mode solitons and vortices. This implies that introducing bright defects (defects with amplitude higher than the surrounding background) may otherwise aid the creation and stabilization of dark gap localized modes. This is indeed true, as shown below.

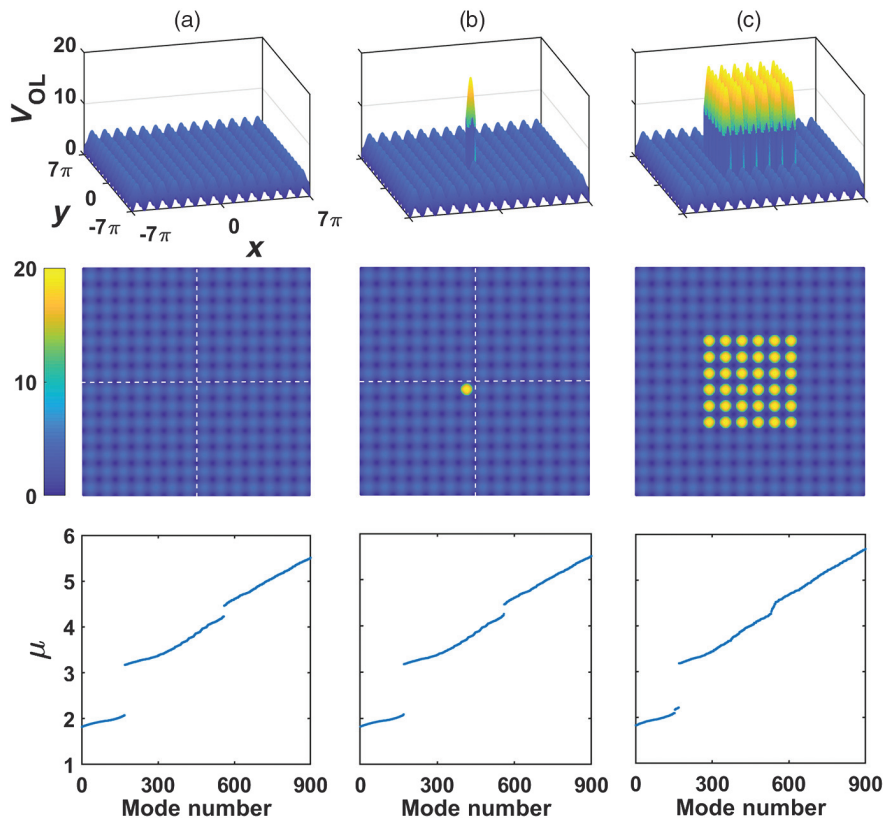

**Fig. 3** Calculated 2-D profiles of the optical periodic potentials ($V_{\text{OL}}$, first row) and their contour plots (central row) as well as the corresponding linear spectra (bottom row): (a) perfect optical lattice, optical lattices with (b) single and (c) multiple bright defects (with the number of defects $n = 36$). Here and below, $V_0 = 2.5$ and $V_0 = 10$ are used, respectively, for optical lattices without and with defects. Perpendicular dashed lines in the central two panels represent two coordinates ($x$ and $y$), with the intersection corresponding to origin of coordinates, point (0, 0).

To investigate whether a particular 2-D defect structure allows for dark soliton formation, the 2-D optical potential structures for the homogeneous optical lattices with $V_0 = 2.5$, and the ones with single and multiple defects with $V_0 = 10$, are separately displayed in Figs. 3(a)–3(c). Their contour plots are displayed in the center row; their linear spectra, calculated directly from the linearized model of Eq. (2), are displayed in the bottom row. It is observed from the bottom row panels that the second BG gradually constricts with an increase in the number of defects, and the same thing also happens for the first BG. We stress that, although the defect potential arranged in Fig. 3(c) is in the same period of the optical lattice, unequal periods for both potentials can also be considered; speculatively, its effect on the research results and conclusions of this paper is relatively low. It is relevant to mention that, in Figs. 3(a) and 3(b), we have plotted two perpendicular dashed lines in the central row panels to illustrate the $x$ and $y$ coordinates under consideration, stressing that their intersection is the coordinate origin (0, 0); this indicates that the 2-D dark gap solutions reported below also station at the minimum of the optical lattice [the same feature as their 1-D cases, e.g., Fig. 1(e)], thus the bright defect is also introduced at the minimum positions of the lattice.

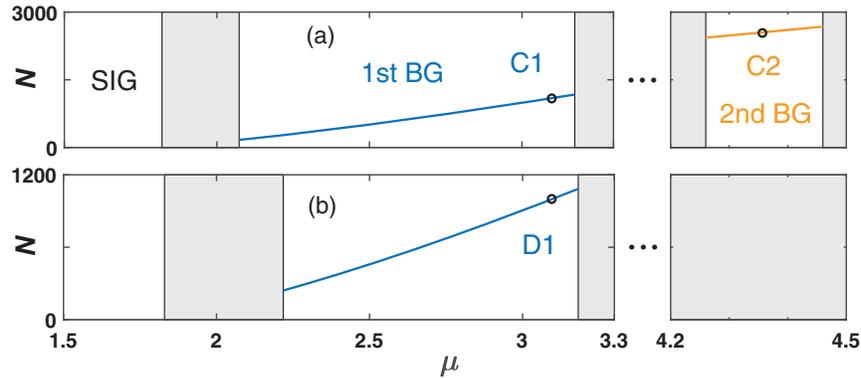

**Fig. 4** Number of atoms ($N$) versus chemical potential $\mu$ for 2-D matter–wave dark localized modes: (a) dark gap solitons with 1 defect and (b) dark gap soliton clusters supported by the 2-D periodic optical potential (optical lattice) with 36 defects. The three-dimensional density distributions, contour plots, and profiles for the marked realizations C1, C2, and D1 are shown in Figs. 5(a)–5(c), respectively.

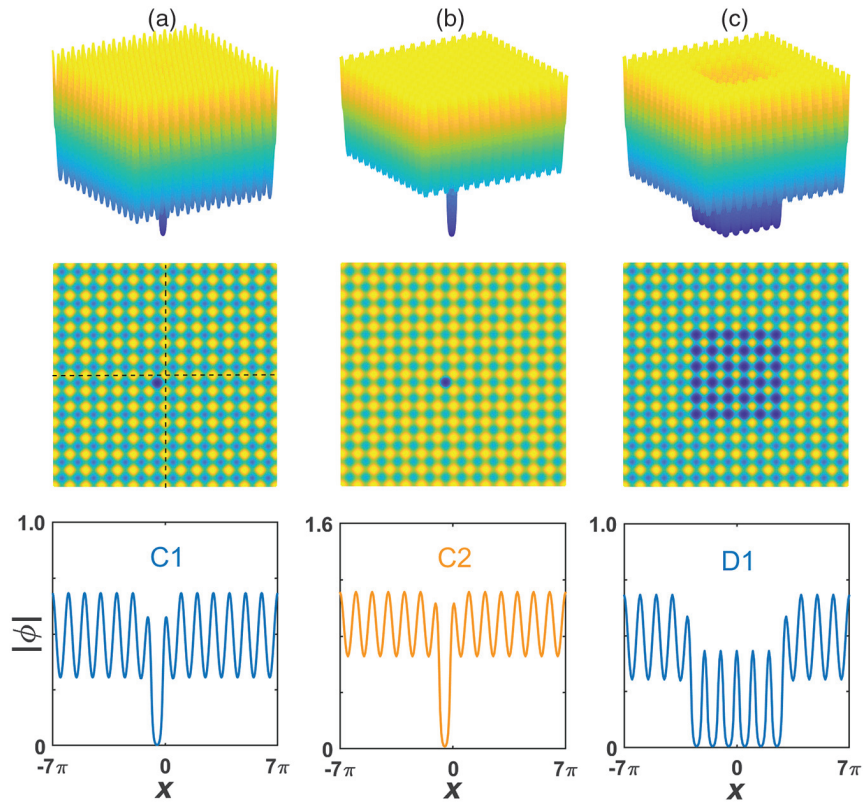

**Fig. 5** Calculated atom density distributions (first row), their contour plots (central row), and the profiles (bottom row) of 2-D matter–wave dark gap modes: dark solitons in (a) the first BG with $\mu = 3.1$ and (b) the second BG with $\mu = 4.36$; (c) dark gap wave (soliton clusters) in the first BG with $\mu = 3.1$. Dashed lines in the first central panel represent the Cartesian co-ordinate system.

Our numerical simulations demonstrate that the 2-D dark gap solitons, built on nonlinear Bloch-waves background and supported by optical lattices with a single defect that leads to the formation of a central dip of the dark mode, can be stable modes, as verified numerically by their relevant dynamical evolution with time $t$, as demonstrated in the Supplementary Material. The dependence $N(\mu)$ for the 2-D fundamental dark gap solitons within first and second BGs is shown in Fig. 4(a); the same curve for dark gap soliton clusters is depicted too in Fig. 4(b). Note that, once again, the anti-VK stability criterion $dN/d\mu > 0$ is obeyed. Comparing both panels, we find that the number of atoms (norm $N$) decreases with an increase of the number of dark solitons, contrary to their bright counterparts. This may be explained by the fact that increasing central dips of dark solitons serves to evaporate some active cooled atoms and thus, diminishes the atom number. Representative examples of such dark gap solitons are shown in Figs. 5(a) and 5(b), with the atom density distributions, contour plots, and profiles. It is seen that the central dip of the 2-D dark gap solitons pinned in a single defect narrows, while the amplitude of the solitons increases, when increasing $\mu$ (e.g., going from the first BG to the second one); this pattern accords with their 1-D counterparts as shown in Fig. 1. The optical lattice with more defects can support a new kind of dark gap solitonic structure: dark soliton clusters whose typical example is shown in Fig. 5(c). Direct numerical simulations further confirm that both types of the 2-D dark localized modes—gap solitons and their composite, soliton clusters—are dynamically stable. The physical explanation for this attribute is that they are the intrinsic fundamental localized defect modes of the underlying model, Eq. (2).

We next focus on the generation of vortical counterparts of the thus-found dark gap solutions, nontrivial phase states. In Fig. 6(a), we give an example of a matter–wave dark gap vortex carrying vorticity (topological charge) $m = 1$, which was generated via phase imprinting with the ground-state nonlinear Bloch wave as background. Numerically, this is done by simply adding the term $\exp(im\theta)$ in the above-predicted soliton structures of both types. Systematic simulations verified that the dark gap vortices are dynamically stable at $m = 1$, while unstable at $m \geq 2$. Stable vortical modes are also for the dark gap soliton clusters; in particular, they are stable under the condition $m \leq 2$, while completely unstable at $m \geq 3$ [see their stable and unstable examples in Figs. 6(b) and 6(c), respectively]. The dynamics of vortical modes of both types [dark gap vortices and soliton clusters carrying vorticity, over time $t$ by monitoring their evolutions in real-time model, Eq. (1)] are presented in the Supplementary Material.

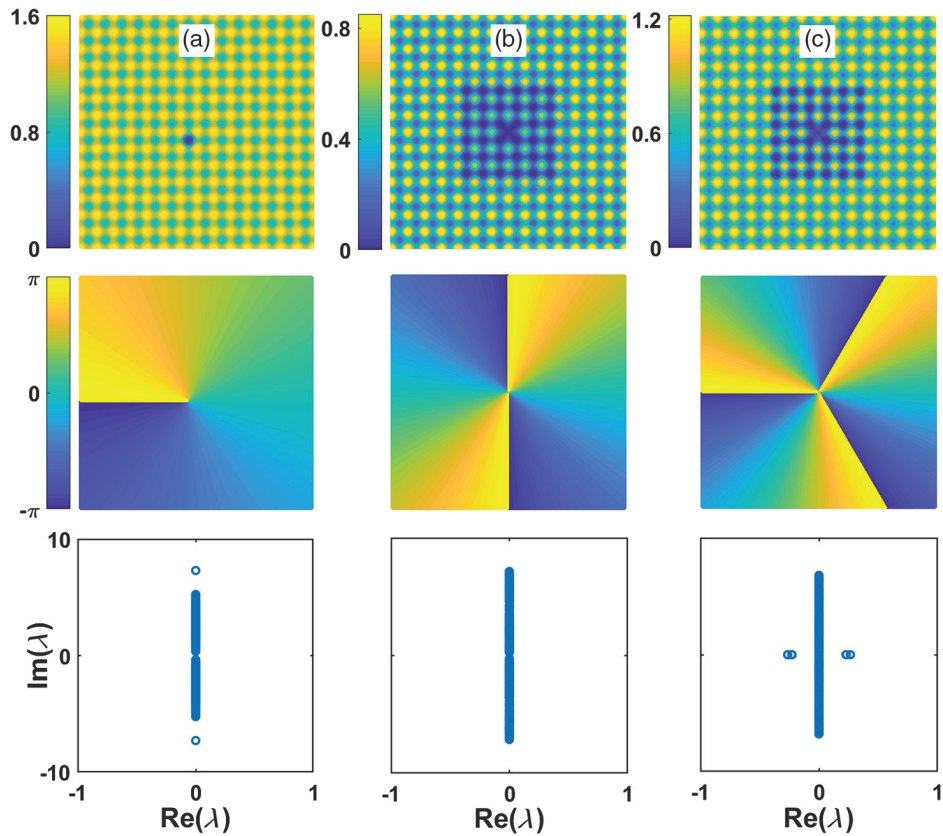

**Fig. 6** Contour plots of the atom density distribution (top), phases (central), and eigenvalues (bottom) of the 2-D matter–wave dark gap modes with engraved vortex: with vortex charge (a) $m = 1$ at $\mu = 4.36$, (b) $m = 2$ at $\mu = 2.4$, and (c) $m = 3$ at $\mu = 3.1$. Panel (a) represents dark gap vortex; panels (b) and (c) represent stable and unstable vortex states of dark gap soliton clusters, respectively.

## 4 Conclusion

It is known that, within the mean-field framework, the creation of gap-type dark localized modes—including dark gap solitons and gap vortices, alike, inside the finite BG of the underlying linear spectrum region caused by placing a BEC on optical lattices in dimensions greater than one—is a challenging issue. To this end, we devised a way to solve the open problem of creating 2-D gap-type dark localized modes, which is done using defect engineering methods used popularly in nonlinear optics[40–44] (see also the Supplementary Material)—inducing single or multiple bright defects to the lattice minimum for optical periodic potential, which enables the formation of stable 2-D dark gap modes appearing as localized defect modes. Note that such bright defects[50–53] are able to be readily fabricated in ultracold atom experiments since, in principle, one can generate the optical lattices with arbitrary configurations,[1–7] and of course include those lattices with arbitrary defects.

Two kinds of dark gap modes—gap solitons and soliton clusters—as well as vortical counterparts of both kinds, were introduced; the existence, dynamics, and stability properties of these modes were explored by means of linear-stability analysis and direct numerical simulations. For the certain physical parameters under consideration, all 2-D fundamental dark structures are stable in the respective gap spectrum. Specifically, the 2-D dark gap solitons can be stable localized defect modes pinned in both the first and second BGs, whereas the dark gap soliton clusters solely exist in the first BG. As far as vortical modes of dark gap solitons and soliton clusters are concerned, stable vortical modes of dark solitons are confined to topological charge (vorticity) $m = 1$, whereas such value can be up to $m = 2$ for the latter case (vortical counterparts of dark gap soliton clusters). The study also covered 1-D dark gap solitons and soliton clusters, although their generation is much different and easier, by simply using the perfect optical lattice (without defect). We stress that, to our knowledge, the dark gap soliton clusters (composed of multiple dark gap solitons) are a new kind of localized wave that has not been explored before. All the dark gap localized modes are centered on the lattice minimum and accompanied by a background that is associated with nonlinear Bloch waves of the corresponding physical model Eq. (1). Stability regions of all the predicted localized solutions in one and two dimensions were given in their relevant parameter spaces. Our theoretical predictions pave the way toward realizing both gap-type dark solitons and vortical ones with ordinary techniques popularly used in current ultracold atom laboratories.[1–9]

A natural extension of this work is to consider a more complicated case—the coupled nonlinear Schrödinger models,[1,54] e.g., two-component BECs imprinted with optical lattices, and clarify the existence and stability of gap-type dark solitons therein. Nowadays, the combined linear-nonlinear lattices models[47,48,55] with periodically modulated linear[49] and nonlinear[56–58] potentials have been researched and widely used to stabilize various species of bright solitons, including gap ones; applying them to the study of dark gap modes should be obvious and interesting. The inhomogeneous nonlinearity distributions,[59] including periodic ones, may be realized via the Feshbach resonance technique.[60] A remaining challenging issue waiting to be solved is how to generate 2-D dark gap modes in homogeneous optical lattices.

We anticipate that the study of dark localized modes structures of gap type, pinned in the finite BGs of the underlying linear spectrum in atomic BEC loaded into optical lattices, will open a new direction to research various other types of excitations, e.g., ring dark solitons[61] and vortex composites,[61] in the same and similar physical settings and beyond (such as in fiber Bragg gratings, layered structures, photonic crystals, and lattices in nonlinear optics),[44,62,63] and stimulate the creative enthusiasm and initiative of scientists to observe the predicted results using modern state-of-the-art technologies that are simple, mature, and easily realizable, and lead the development of such a direction.

After submission of this paper, one of the anonymous reviewers suggested that we survey the existence of 1-D dark gap solitons (and the cluster ones) in the 1-D optical lattices with one or several defects. We have checked this angle in detail and found that such dark gap solitons indeed can be stable localized modes; their existence can be explained by the same physical mechanism as the 2-D cases presented in Fig. 5. An unique property of the 1-D dark gap solitons in between the 1-D perfect optical lattice and imperfect ones with one or few defects is that the dark gap modes supported by the former are on-site modes, whereas the off-site modes are for the latter forms of lattices, with the dips (of the dark modes) resting on the lattice minimum and maximum, respectively.

*Acknowledgments*

This work was supported, in part, by the National Natural Science Foundation of China (Project Nos. 61690224 and 61690222) and the Youth Innovation Promotion Association of the Chinese Academy of Sciences (Project No. 2016357).

**Liangwei Zeng** is a PhD candidate at the Xi'an Institute of Optics and Precision Mechanics of Chinese Academy of Sciences (CAS) and University of Chinese Academy of Sciences. He received BS degree in optical information science and technology from the South China Agricultural University in 2013 and his MS degree in optics from the South China Normal University in 2016. His current research interests include nonlinear optics, Bose–Einstein condensates, and optical solitons.

**Jianhua Zeng** joined the State Key Laboratory of Transient Optics and Photonics, Xi'an Institute of Optics and Precision Mechanics of CAS as an associate professor in July 2013 and became a PhD supervisor after two years. He received his PhD in optics in 2010, jointly trained at Sun Yat-sen University (Guangzhou) and Weizmann Institute of Sciences, Israel, and then did two postdocs at Tel-Aviv University and Tsinghua University, Beijing. His primary research interests focus on theoretical nonlinear photonics, including nonlinear/ultrafast/quantum optics and ultracold quantum gases.